# On the utility of toy models for theories of consciousness

Larissa Albantakis[1,*]

[1] Department of Psychiatry, University of Wisconsin, Madison, WI, USA
* albantakis@wisc.edu

## Abstract

Toy models are highly idealized and deliberately simplified models that retain only the essential features of a system in order to explore specific theoretical questions. Long used in physics and other sciences, they have recently begun to play a more visible role in consciousness research. This chapter examines the potential utility of toy models for developing and evaluating scientific theories of consciousness in terms of their ability to clarify theoretical frameworks, test assumptions, and illuminate philosophical challenges. Drawing primarily on examples from Integrated Information Theory (IIT) and Global Workspace Theory (GWT), I show how these simplified systems could make abstract concepts more tangible, enabling researchers to probe the coherence, consistency, and implications of competing frameworks. In addition to supporting theory development, toy models can also address specific features of experience, as exemplified by the account of spatial extendedness and temporal flow provided by integrated information theory (IIT) and recent theory-independent structural approaches. Moreover, toy models bring philosophical debates into sharper focus, such as the distinction between functional and structural theories of consciousness. By bridging abstract claims and empirical inquiry, toy models provide essential insights into the challenges of building comprehensive theories of consciousness.

## Introduction

Toy models have long been a foundational tool in scientific inquiry, especially in fields where full-scale models or direct experimentation are either impractical or impossible. A toy model is a simplified version of a more complex system that retains only the essential elements necessary to explore specific theoretical questions. These models are deliberately minimalistic, enabling scientists to focus on the core features of an issue without being distracted by extraneous details. In short, toy models are highly idealized and extremely simple models of natural and social phenomena (Reutlinger et al. 2018).

Historically, toy models have been widely used in physics. One of the most famous advocates for using simple, conceptual models to gain insights into complex physical systems was Richard Feynman, who also left us with the memorable words "what I cannot create, I do not understand." His work in quantum electrodynamics, for example, introduced simplified representations of particle interactions to make highly abstract equations more intuitive and tractable. These "Feynman diagrams" are graphical tools that depict how particles like electrons and photons interact. Although these diagrams are not literal pictures of physical processes, they serve as toy models that help physicists visualize and calculate the probabilities of quantum events in a manageable, rule-based way. Another well-known example is the Ising model in statistical mechanics, which is used to study how local interactions among particles can give rise to large-scale phenomena like phase transitions in magnetism. Each particle is represented as a point on a lattice that can be in one of two binary states (spin up or spin down) and interacts only with its immediate neighbors. Despite this simplification, the Ising model captures key features of magnetic behavior, such as the emergence of order and the sudden loss of magnetization at critical temperatures, offering valuable insights into real-world materials. In economics, toy models such as the "rational actor" model or simplified market simulations allow researchers to isolate particular economic drivers, avoiding the complexities of real-world markets. Similarly, in ecology, simple predator-prey models, described by the Lotka-Volterra equations, help biologists understand how the population of two interacting species such as foxes and rabbits can fluctuate over time, without needing to account for the many variables present in a real ecosystem.[1]

Toy models can clarify underlying principles, instantiate minimal mechanisms, foster deeper understanding of complex system, and generate novel predictions about the phenomena under investigation. Unlike complex mathematical analyses, which can rigorously prove functional relationships or implications but often remain abstract and difficult to interpret, toy models implement ideas in a tangible way. They offer a way to explore the mechanics of a system while preserving both clarity and interpretability. Although simplification is not always ideal (think of the "spherical cow" joke in physics[2]), toy models provide an intuitive grasp of processes that would otherwise be challenging to study in their full complexity. In this sense, toy models occupy a middle ground between highly realistic models—often requiring computational simulations that become epistemically opaque black boxes—and purely conceptual thought experiments, which typically lack even minimal mechanistic structure. Additionally, toy models can help identify the minimal conditions required for certain behaviors or outcomes and lead to novel predictions that can be empirically tested. In fields like physics, biology, and economics, toy models are thus not just tools for explanation but also for discovery.[3]

Despite their success in the natural and social sciences, toy models are often viewed with skepticism in consciousness science. One common concern is that consciousness appears to be tightly linked to complexity (Sarasso et al. 2021). Toy models are simple by design and thus lack the very complexity that many believe is necessary for consciousness to emerge (e.g., Aaronson 2014). Another concerns relates to the dynamical role of consciousness: if a system's behavior can be fully understood and predicted from its physical dynamics, there may seem to be no explanatory role left for consciousness (Leibniz 1714; Kleiner and Ludwig 2023). If one takes dynamical relevance to be a necessary condition for consciousness, then a fully specified toy model would be excluded as a plausible substrate. Nevertheless, when reduced to their essential features, the mechanisms proposed by various neuroscientific theories to underlie conscious experience can often be implemented using only a small number of interacting neurons. This has been referred to as the "small network argument" (M. H. Herzog et al. 2007; Doerig et al. 2020), and highlights another challenge: due to the inherently subjective nature of consciousness, we cannot assess whether a toy model that meets the essential criteria of a given theory is genuinely conscious in the same way we might validate predictions in other scientific domains. Theories of consciousness can only be validated through our own experience (Albantakis 2020; Tononi et al. 2025; Tononi 2014), in healthy adult humans who can report on their experiences. Given these limitations, what purpose can toy models serve in consciousness science?

---

[1] As these examples illustrate, toy models can take many forms. Reutlinger et al. (2018) distinguish two broad classes: *embedded* and *autonomous* toy models. Embedded toy models are simple and idealized models of phenomena developed within an established framework theory. In contrast, autonomous toy models are largely independent of any particular theory. For example, the Ising model is an embedded toy model in statistical mechanics, used to demonstrate how local interactions can give rise to phase transitions. By contrast, the Lotka-Volterra predator-prey model is an autonomous toy model: it helps to elucidate plausible population dynamics without presupposing a comprehensive theory of ecology. The focus of this chapter is largely on toy models embedded within theories of consciousness, developed to illustrate, refine, or evaluate theoretical claims. Nevertheless, autonomous toy models are also beginning to emerge in the study of consciousness, aiming to explore the structural features of subjective experience itself (e.g., Prentner 2019).

[2] The spherical cow is a humorous metaphor for overly simplified models of complex phenomena that allow for mathematical expression but no longer reflect reality in the relevant sense.

[3] Throughout, I reserve the term *model*—and specifically *toy model*—to refer to tools for exploring theoretical claims (e.g., simplified simulations, concrete idealized examples, or instantiations). I use *theory* for explanatory frameworks, even though it has been argued that many so-called theories of consciousness might more accurately be described as models themselves (Signorelli et al. 2025).

In the following, I advocate for the utility of toy models in consciousness science, particularly in the development of principled theories of consciousness. First, toy models can be used to evaluate the coherence and explanatory power of theories of consciousness, ensuring they are more than abstract ideas removed from physical implementation. Second, toy models can clarify the predictions and implications of theories of consciousness. By providing concrete examples, toy models can bridge the gap between abstract theoretical claims and empirical tests, guiding experimental research and helping refine theories based on internal and empirical consistency. Finally, toy models can help bring philosophical challenges to light, exposing methodological limitations of prevailing approaches and points of contention in the field.

## Theory-driven approaches to consciousness and "good" explanations

The contemporary science of consciousness initially focused on identifying the neural correlates of consciousness (NCCs), a deliberately "theory-neutral" approach aimed at characterizing the presence or absence of consciousness, as well as specific contents of consciousness, in human subjects. NCCs are typically defined as the minimal neural mechanisms jointly sufficient for a specific conscious experience. However, as consciousness science has progressed, the limitations of this approach have become clear, particularly in its inability to generalize beyond human-like neural systems and its failure to provide a principled account of what consciousness *is* in physical terms (understood here as a systematic correspondence between conscious states and physical systems, without metaphysical assumptions). While some set of neural mechanisms may be sufficient to reliably infer conscious experience in healthy adult humans, this does not help us assess consciousness in structurally or functionally different systems, such as infants, non-human animals, or intelligent machines, which may lack the same mechanisms but could still support a similar experience in different ways. These limitations, along with the rapid rise of artificial intelligence, have prompted a shift in focus toward theories that aim to account for the nature of consciousness and offer principled frameworks that are generalizable beyond biological brains. While "theory-neutral" or "theory-light" approaches (Birch 2022) may have some merit when extrapolating from humans to animals, where we can rely on evolutionary relatedness and similarities in cognitive function, drawing on principled theories becomes essential when evaluating consciousness in artificial intelligence (Butlin et al. 2023; Findlay et al. 2024).

This growing emphasis on theory-driven approaches has led to a proliferation of candidate "theories." Illustrative of this trend is Kuhn's (2024) recent attempt at an extensive taxonomy of the "landscape of consciousness," surveying proposed explanations spanning from philosophical accounts to quantum physics. A more focused overview is offered by Signorelli et al. (2021), who provide a systematic classification of the explanatory profiles of different theories of consciousness. Focusing solely on neurobiological theories of consciousness, Seth and Bayne (2022) counted 22 current proposals, noting that, rather than being progressively ruled out as empirical data accumulate, proposed hypotheses about consciousness seem to be multiplying.

One challenge with many of these ideas is that they are formulated either in abstract (i.e., mechanistically vague) terms, or have limited scope, primarily focusing on neural mechanisms related to consciousness in humans and closely related species. While many approaches propose—or can be interpreted to propose (Shevlin 2021)—general principles that should, in theory, be applicable to any physical system, the mechanistic details provided are usually insufficient for constructing a formal framework that would allow evaluating whether a given physical system is conscious and in what way (Kanai and Fujisawa 2024). This lack of specificity can make it difficult to assess whether a theory proposal is coherent and has merit as a "good"[4] explanation for consciousness—one that can account for a broad set of facts (*scope*), does so in a unified manner (*synthesis*), explains facts precisely (*specificity*), is internally coherent (*self-consistency*), aligns with our broader understanding (*system consistency*), is simpler than alternatives (*simplicity*), and can make testable predictions (*scientific validation*) (Albantakis, Barbosa, et al. 2023).

Toy models in this context may facilitate a deeper understanding of the theories that aim to explain it, benefitting both skeptics and proponents alike. Moreover, differences in predictions among current accounts of consciousness suggest inconsistencies that may arise not solely from the diversity of proposals, but from more fundamental issues, such as mechanistic incoherence or incompatibility with physical principles (Kleiner and Hartmann 2023). Even in the absence of decisive new evidence from human subjects, toy models offer a way to instantiate abstract concepts in simplified systems, providing a platform to clarify and scrutinize the theoretical assumptions and conditions each theory claims are essential for conscious experience. For example, how does a global workspace posited by Global Workspace Theory (GWT) actually look like when we try to construct one? Which types of neural architectures genuinely support a high degree of integrated information as required by Integrated Information Theory (IIT)? Is there really a qualitative difference between top-down predictions and bottom-up prediction-error signals, given their distinct phenomenal roles in predictive processing accounts of consciousness? And, most critically, is a given theory coherent in the first place? In this sense, toy models may serve as testing grounds for assessing the explanatory power, mechanistic plausibility, and coherence of theories of consciousness.

---

[4] The quotation marks around "good" are meant to indicate a modest and pragmatic use of the term, adapting the common notion of inferences to the *best* explanation, as in (Albantakis, Barbosa, et al. 2023).

Furthermore, toy models provide a framework to relate different theories of consciousness to each other. An initial attempt can be found in (Lundbak Olesen et al. 2023), which applied both a measure of integrated information and surprisal (a basic information theoretic measure from the free energy principle formalism) to small, artificial agents evolving in a simulated environment.

## Two case studies—IIT and GWT

Despite a proliferation of proposed hypotheses, consciousness science has coalesced around a few prominent theoretical frameworks, each with distinct explanatory goals and methodologies (Yaron et al. 2021). Among these, global workspace theory (GWT) and integrated information theory (IIT) are two of the most influential, along with higher-order theories (HOTs), re-entry theories, and predictive processing theories (Seth and Bayne 2022). Not only do these frameworks propose different explanations for consciousness, but they also target different aspects of conscious experience (Signorelli, Szczotka, et al. 2021). While IIT centers on phenomenal features of consciousness, aiming to explain the subjective quality of experience, others, such as GWT, emphasize functional, behavioral aspects. In the following, I will focus specifically on IIT and GWT, as they represent two contrasting approaches to consciousness—one grounded in phenomenology[5], the other in functional cognitive architecture—and examine the role of toy models in evaluating these frameworks.

### Integrated Information theory (IIT)

Most neurobiological theories of consciousness start from experimental observations, aiming to explain specific phenomena and eventually derive general principles. In contrast, IIT has strived from the outset to provide a principled and comprehensive account of consciousness. To that end, IIT starts from consciousness itself (as opposed to potential physical correlates), aiming to identify the essential properties of conscious experience through introspection and reasoning, and to translate them into postulates about its physical substrate. The resulting framework can then be evaluated against experimental data, with the ultimate goal of creating an objective account of the presence and properties of experience (Albantakis, Barbosa, et al. 2023). IIT concludes that a substrate of consciousness, constituted of interacting units, must be a maximum of irreducible cause-effect power. Furthermore, IIT proposes an explanatory identity between the cause-effect structure supported by such a "complex" in its current state and the quality of its experience (see IIT Wiki (2024) for more information and Bayne (2018) and McQueen (McQueen 2019) for a critical perspective on IIT's axiomatic approach).

From its inception, IIT's conceptual claims have been accompanied by a developing mathematical framework. Each successive refinement and publication has featured toy models that clarify and illustrate the proposed formalism and its implications (Tononi et al. 1994; Tononi and Sporns 2003; Balduzzi and Tononi 2008; Oizumi et al. 2014; Albantakis, Barbosa, et al. 2023). A typical IIT toy model consists of a small network constituted of 3 to 6 binary units ("toy neurons") that interact according to predefined update rules. These units may be implemented as simple logic gates (e.g., AND, OR, XOR) or as probabilistic elements governed by state-transition probability functions. Such simple systems allow for a rigorous evaluation of the causal and informational quantities specified by the IIT formalism, which makes it possible to explore how architectural and functional features influence the quantity and quality of experience according to the theory.

---

[5] Here and throughout, "phenomenology" refers to the structure and character of subjective experience itself, not the philosophical tradition of Phenomenology associated with Husserl.

These toy models have not only been used to elucidate IIT's theoretical framework in publications but have also served as test cases for assessing its internal coherence. They have driven advancements such as a novel measure of intrinsic information (Barbosa et al. 2020), an account of causal emergence and macroscopic intrinsic units, with toy models providing proof-of-principle examples (Hoel et al. 2013; 2016; Marshall et al. 2018; 2024), and formal extensions to quantum computational systems (Zanardi et al. 2018; Albantakis, Prentner, et al. 2023).

While, on the one hand, the use of toy models has made the IIT framework sufficiently specific and accessible to invite criticism and opposing views (e.g., (Aaronson 2014; Hanson and Walker 2019; Doerig et al. 2019; Merker et al. 2021), on the other hand, it served to illustrate its explanatory and predictive power. For example, IIT explains why the cerebellum, despite having more neurons than the cortex and being connected to the rest of the brain, does not contribute to experience due to its modular and primarily feedforward anatomy. It also accounts for why consciousness is lost when the causal interactions among cortical neurons break down, such as during deep sleep (when neurons become bistable) or seizures (when neural activity becomes saturated and unresponsive). Simple examples also suffice to demonstrate that, according to IIT, silent neurons with functional connections may contribute to experience (Albantakis, Barbosa, et al. 2023)—a prediction currently under evaluation (Olcese et al. 2024, Experiment 1). Similarly, the prediction that changes in connectivity within the main complex should lead to changes in experience even without changes in activity can be captured by simple toy examples and tested in human subjects (Tononi et al. 2016; Song et al. 2017).

In part, the prominent role of toy models in IIT arises from necessity: rigorous application of the mathematical framework is feasible only in systems with small numbers of discrete units. This is because the state space, the number of subsystems, and the system partitions that must be evaluated all grow exponentially with system size, rendering the computations intractable for anything but very small networks. Consequently, some aspects of IIT are more readily testable than others, which has prompted a distinction in the secondary literature between "weak" IIT, focused on empirical correlates of integrated information, and "strong" IIT, which aims to provide a universal account of consciousness that identifies experiences with the cause-effect structures of maximally irreducible substrates (Mediano et al. 2022; Leung and Tsuchiya 2023). While the empirical testability of "strong" IIT has been challenged (Michel and Lau 2020; Klincewicz et al. 2025; but see (Tononi et al. 2025), toy models offer a way to connect more specific theoretical implications of the theory in the strong sense with empirically accessible observables. As in other disciplines, insights drawn from simple examples representing different types of neuronal architectures (e.g., modular, grid-like, random, or all-to-all) or mechanisms (e.g., linear, nonlinear, excitatory, or inhibitory) can be generalized to larger systems. These generalizations enable IIT to provide explanations and predictions that can be validated in human subjects.

Toy models in IIT have also been used to elucidate specific phenomenal experiences, going beyond their use-case of illustrating the theoretical framework itself. For instance, IIT explains the experience of spatial extendedness using non-directed grids, whose cause-effect structures are composed of relations arranged according to reflexivity, inclusion, connection, and fusion—mirroring spatial phenomenology (Haun and Tononi 2019). Similarly, the feeling of temporal flow can be accounted for by the cause-effect sub-structures specified by arrays of directed grids (Comolatti et al. 2024). These highly specific predictions address core principles of IIT and demonstrate its capacity to link theoretical postulates with concrete phenomenological accounts. Notably, the structural features of consciousness have recently gained renewed attention beyond the context of IIT (Kleiner 2024). Within the emerging framework of mathematical consciousness science (Kleiner and Ludwig 2024; Signorelli et al. 2025; Prentner 2024), several studies have introduced toy models that are not embedded within a specific theory of consciousness but instead aim to directly characterize or explain particular aspects of experience such as its unity, its compositional character, and its subjectivity (Prentner 2019; Signorelli, Wang, and Khan 2021; Signorelli, Wang, and Coecke 2021; Mason 2021; Díaz-Boils et al. 2025). These autonomous toy models prioritize the internal structure of consciousness itself and exemplify an alternative use case: exploring experience from a structural standpoint rather than deriving it from a broader mechanistic theory.

Returning to IIT, toy networks have also been used to explore broader questions about the evolution of consciousness and the possibility of consciousness in artificial systems. Simulated evolution experiments with simple agents equipped with evolvable neural networks have shown that integrated information increases over generations when agents face selective pressures under biological constraints in sufficiently complex environments (Edlund et al. 2011; Albantakis et al. 2014; Fischer et al. 2020). These findings provide a possible explanation, grounded in IIT, for why complex conscious systems evolved and why consciousness and intelligence correlate in biological systems, even though they can be dissociated in principle, meaning that a system may exhibit intelligent behavior without being conscious. This dissociation has been demonstrated using a toy-model implementation of a standard computer (Findlay et al. 2024), which exposed that computers do not typically specify cause-effect structures that resemble those of the systems they simulate. According to IIT, this means that functional equivalence does not imply phenomenal equivalence, which challenges widely held computational-functionalist assumptions (Butlin et al. 2023).

### Global workspace theory (GWT)

In contrast to IIT's "phenomenology-first" approach, GWT originated in cognitive science, drawing inspiration from artificial intelligence research on cognitive architectures (Baars 1988). It was subsequently developed into a neurobiological model known as the Global Neuronal Workspace Theory (GNWT) (Dehaene et al. 1998; 2003; Dehaene and Changeux 2011). GNWT posits that (sensory) information becomes conscious when it enters and is "broadcast" within an anatomically widespread neural workspace, primarily involving higher-order cortical association areas, with a particular (though not exclusive) emphasis on the prefrontal cortex (Mashour et al. 2020; Seth and Bayne 2022).

Computational models have been central to GWT and GNWT, but they function primarily as simplified models of neurobiological processes rather than as toy models aimed at evaluating GWT as a theory of consciousness. Early conceptual sketches of cognitive architectures outlined the minimal components thought to be required for consciousness—distributed specialized processors and a reciprocally connected global workspace or 'blackboard'—in abstract terms, without mechanistic detail (Baars 1988). In contrast, mechanistic models have been used to target specific neural phenomena, such as an amplification of perceptual activity ("ignition"), long-distance correlations, or the P300 waveform, while explicitly denying that these are exhaustive models of conscious substrates (Dehaene et al. 1998; Dehaene and Naccache 2001). Although these models provide testable hypothesis about psychological phenomena such as the attentional blink and inattentional blindness (Dehaene et al. 2003; Dehaene and Changeux 2005), they are not aimed at exploring minimally sufficient mechanisms or clarifying the core principles of GWT/GNWT as a theory of consciousness.

As with other neurobiological theories of consciousness, but in contrast to IIT, the scope and generality of GWT/GNWT remain, to some extent, open to interpretation (Birch 2022; Shevlin 2021). Specifically, it is unclear whether the computational principles underlying GWT are generally sufficient for conscious experience, or if the specific neural mechanisms proposed by GNWT in healthy, adult humans are required to make meaningful predictions about the presence or absence of consciousness in a given system. As suggested in (Doerig et al. 2020), a network consisting of two peripheral neurons connected to a small recurrent global workspace meets the functional criteria for consciousness outlined by GWT. Yet, according to (Dehaene et al. 1998) even their simple neural model of a global workspace was explicitly not intended to provide "an exhaustive description of a 'conscious workspace.'" This leaves open the question of what additional features, if any, might be required to construct a conscious system under this framework, an issue discussed further below.

A "cautious" interpretation of GWT merely implies that the presence of a global broadcast network in healthy adults is sufficient for consciousness (Birch 2022). While this avoids potential overgeneralization, it has no bearing on the presence or absence of consciousness in systems different from us and does not provide an account of the nature of consciousness. By contrast, an "ambitious" interpretation of GWT would imply that the presence of specialized modules competing for access to a global workspace, combined with the capacity for global broadcast, could be taken as evidence for consciousness (Birch 2022). For instance, Dehaene, Lau, and Kouider (2017) proposed a functional definition of consciousness that could, in principle, extend to machines, while leaving open the question of phenomenal consciousness. However, the lack of a principled formal framework to rigorously determine whether an arbitrary system possesses a global workspace—or to precisely define what constitutes broadcasting—limits the applicability of this interpretation (Seth and Bayne 2022; Birch 2022; Kanai and Fujisawa 2024).

Since GWT is generally presented as a computational functionalist theory of consciousness, various attempts have been made to formalize its principles in computational terms (Franklin and Graesser 1999; VanRullen and Kanai 2021; Goyal et al. 2022). In the spirit of a true toy model, Blum and Blum (2021) introduce the Conscious Turing Machine (CTM) as a formalization of GWT, prioritizing simplicity over complexity to provide a minimal model of a conscious system rather than a detailed simulation of the brain. The CTM was proposed for the express purpose of understanding Baars' Theater Model and for providing a theoretical computer science framework to understand consciousness. Central to the CTM is an expressive "inner language" called Brainish, which facilitates global broadcast and enables inner speech, vision, and sensations. This inner language is considered a key component of the feeling of consciousness, though its articulation remains to be developed. Blum and Blum's proposal is notable in its explicit aim to address the minimal requirements for phenomenal experience, distinguishing these from mere simulations of such experiences. However, the precise nature of the "feeling of consciousness" remains unspecified, and the authors point to IIT as a potential framework for addressing this challenge.

In sum, existing simple models of GWT provide basic instances of functional workspaces and serve to elucidate the architectures and dynamics of global workspaces. However, they also reveal an open challenge for GWT/GNWT: the absence of a clearly articulated set of necessary and sufficient conditions for a system to be conscious within the current framework. This vagueness is not necessarily problematic within a functionalist framework of consciousness, where the focus is on utility and behavioral outcomes. Indeed, general systems can be assessed against broad criteria like those proposed by Butlin et al., (2023) who compiled a list of indicators of phenomenal consciousness for assessing AI consciousness based on an assumption of computational functionalism. Nevertheless, many systems ranging from the minimal examples above to real-world cases like infants or patients with severe brain would occupy a gray zone in which existing criteria are insufficient to establish the presence or absence of experience. While the richness of experience of such systems could vary widely, subjectively, it either feels like something to be that system or it does not. As Kanai and Fujisawa (2024) put it: "Concepts such as 'global workspace' (…) are understood by human neuroscientists. However, deciding whether they are present in an arbitrary physical system requires more precise mathematical definitions to allow their identification."

## Philosophical challenges brought to light

Beyond their role as tools for elucidating theoretical frameworks, toy models have been instrumental in clarifying and addressing key philosophical challenges in consciousness science. By providing simplified yet tangible instantiations of theoretical claims, they expose methodological limitations and sharpen debates on foundational issues. For instance, toy models have focused discussions about the minimal conditions for consciousness, baring the choice between postulating additional, often arbitrary requirements to exclude simple systems or accepting that many theories imply consciousness in very basic systems. Similarly, toy models have underscored the tensions between computational functionalist and structural theories of consciousness, exposing key differences and shedding light on the assumptions underlying these competing perspectives.

### Minimal conscious systems

Ultimately, any scientific theory of consciousness that aims to be comprehensive should provide a principled account of what consciousness *is* in physical terms, necessary and sufficient conditions for the presence or absence of consciousness that can be evaluated in a wide range of physical systems, and an account of why an experience feels the way it feels (Ellia et al. 2021). Notably, a principled account necessarily implies that there is a minimal system consistent with the theory that should be granted consciousness—if minimally so[6].

---

[6] With "minimal system" I refer to the simplest (type of) physical system that satisfies a theory's conditions for consciousness. Such a system can be instantiated as a concrete, mechanistically interpretable toy model to explore

These minimal systems challenge our intuitions about consciousness, which often link it to intelligence, likely because in biological systems, the two seem closely connected. As a result, toy model instantiations of such minimal systems provoke discomfort, prompting assertions that the proposed features may be necessary but cannot be sufficient for consciousness. For example, proponents of GWT generally resist attributing consciousness to toy models of global workspaces, implying that additional criteria are needed to explain why these systems are not conscious (Doerig et al. 2020). However, adding arbitrary requirements, such as a minimal size, complexity threshold, or restriction to biological substrates, without offering additional explanatory power, is unscientific if done merely to avoid an uncomfortable implication.

Alternatively, one could accept that any system implementing all the essential features postulated by a theory of consciousness would indeed be (minimally) conscious (Lamme 2006; Albantakis, Barbosa, et al. 2023). This, however, raises concerns among critics who worry that such implications could lead down a slippery slope toward panpsychism— a philosophical position often criticized for its lack of empirical testability and its perceived risk of undermining the scientific utility of the concept of consciousness (Doerig et al. 2020; Merker et al. 2021; Seth 2021; Klincewicz et al. 2025). Panpsychism holds that consciousness is a fundamental and ubiquitous feature of reality, which, if taken literally, suggests that some form of consciousness is present in all things. The concern is that if a theory allows for simple systems to be conscious, it risks failing to address the rich and complex features of human conscious experience. This concern is particularly relevant for theories like GWT, which offer broad functional explanations but do not address the qualitative aspects of experience. Such theories inherently struggle to differentiate between minimal and complex systems. If consciousness arises from the global broadcast of information, what distinguishes a small global workspace from a larger one? Moreover, without a principle to arbitrate between nested or overlapping workspaces, GWT cannot resolve the ambiguity of whether subsets (or supersets) of a larger workspace that meet its functional criteria might also independently qualify as conscious. Without a framework to explain how the scale, structure, or extent of the global workspace shapes the quality and content of experience, these theories leave unanswered the fundamental questions of why and how the rich and complex conscious experiences of humans differ from the simpler, potentially conscious states of minimal systems operating under equivalent functional principles.

These unresolved issues highlight the great challenge faced by any comprehensive theory of consciousness: it must account for every aspect of experience, from its essential properties, such as the unity of consciousness, to its accidental properties, like the fleeting feeling of familiarity evoked by a faintly remembered tune. While most proposed theories of consciousness are narrower in scope, IIT stands out in its aim to provide a comprehensive account. According to IIT, a substrate of consciousness must specify a maximum of system integrated information, a quantity defined to reflect the essential properties of experience identified by IIT in physical terms. In addition, every accidental property of experience—the feeling of spatial extendedness, of temporal flow, of objects, of colors, and so on—must be fully accounted for by the properties of the substrate's cause–effect structure, with no additional ingredients. While IIT is often associated with panpsychism (Merker et al. 2021; Tononi and Koch

---

a theory's implications. I do *not* mean minimal models of explanation in the sense of autonomous, idealized representations that aim to capture generalizable abstract features of consciousness.

2015; Klincewicz et al. 2025), it is far from attributing consciousness indiscriminately to all things (Kanai and Fujisawa 2024; Tononi et al. 2025). Whether its exacting predictions can be empirically validated remains an open question. Crucially, though, the universality of IIT does not prevent it from making precise, testable predictions about the causal structure underlying complex human experiences. Toy models are instrumental in formulating these predictions, as they enable a full evaluation of cause–effect structures in systems that are explicitly defined and analytically tractable.

Whether IIT proves to be correct or misguided, any complete theory of consciousness will eventually require us to acknowledge that its proposed criteria are sufficient conditions for attributing consciousness to any system that satisfies them, no matter how simple or counterintuitive that system might seem. As Kuhn aptly stated, "any theory of consciousness, to be complete and sufficient, must make an identity claim. (…) Something happening or existing in every sentient creature just *is* consciousness." A comprehensive scientific theory of consciousness must, in some way, connect subjective experience to the natural world and should thus offer fundamental principles that account for conscious experience in any system. Toy models provide a clear and principled way to explore the coherence of these principles, bridging the gap between abstract theory and empirical validation. A theory that merely predicts which regions of the primate brain correlate with consciousness will fall short of addressing "why," "how," and under which conditions consciousness arises from a physical entity (Kanai and Fujisawa 2024).

### The structural-functional divide

As demonstrated by a series of simple example systems, IIT allows for functional equivalence without phenomenal equivalence (Albantakis and Tononi 2019; Albantakis, Barbosa, et al. 2023; Hanson and Walker 2019; 2020). Specifically, two systems may exhibit identical input-output functions (probing a subset of possible system states), or even share the same internal global dynamics (whole-system state transitions), yet differ in their amount of system integrated information and intrinsic cause-effect structures. This is because identical global dynamics and input-output behavior can arise from different physical systems with distinct internal causal structures, meaning that their components interact in very different ways. For IIT, the quality of experience corresponds to the unfolded cause-effect structure of a substrate, which captures the irreducible cause-effect information of every subset of the substrate, not just the system as a whole. In other words, whether and how a system is conscious depends on what the system is, in causal terms, rather than what it does. This implication challenges the dominant computational-functionalist paradigm, which holds that performing computations of the right kind is both necessary and sufficient for consciousness (Butlin et al. 2023).

Since these implications were made explicit through a set of toy examples, they have sparked an ongoing philosophical debate—primarily targeting IIT—about the testability of theories that put constraints on the types of physical systems that could serve as substrates of consciousness, which also include Recurrent Processing Theory (RPT) and others (Doerig et al. 2020). Introduced as the "unfolding argument"(Doerig et al. 2019) and generalized to the "substitution argument" (Kleiner and Hoel 2020), the central claim is that if it is possible, even in principle, to replace the mechanisms of a conscious system in a way that renders it unconscious without changing its outward responses, then those mechanisms cannot be necessary to explain empirical data about consciousness. The premises underlying the unfolding argument and its conclusions have since been challenged on multiple fronts, including practical, methodological, and philosophical concerns (Usher 2021; Tsuchiya et al. 2019; Negro 2020). One particularly relevant issue is the argument's refusal to recognize first-person experience as a means to validate reports in healthy adult humans—validation that would be absent in the substituted system (Albantakis 2020).

While IIT is unapologetically structural in its approach, whether other neurobiological theories imply specific causal implementations is a matter of interpretation. For instance, Butlin et al. (2023), include RPT in their list of theories compatible with computational functionalism, provided the requirement for recurrent processing is interpreted algorithmically rather than as necessitating a specific causal implementation. Similarly, a global workspace architecture can be understood as either a purely functional construct—implementable by something as abstract as a giant input-output lookup table (e.g. Herzog et al. 2021), or as implying a more constrained implementation (Blum and Blum 2021). However, a purely functionalist interpretation would demand principled methods to determine which input-output functions genuinely imply global workspaces—methods that have yet to be developed. For instance, what specific behaviors would demonstrate that a lookup table truly implements a global workspace?

The question of whether functional equivalence necessarily implies phenomenal equivalence has gained new urgency in the era of advanced artificial intelligence. In this context, IIT demonstrates through a toy model of a standard computer simulating a simple integrated system that the cause-effect structure of the computer diverges fundamentally from that of the system it simulates (Findlay et al. 2024). In contrast, computational functionalism asserts that phenomenological and functional equivalence are inherently linked. However, the resolution of this debate will not depend on the increasing sophistication of artificial systems but instead on the explanatory power of each theoretical framework when applied to human consciousness. As argued earlier, a comprehensive theory of consciousness must account for an astonishing breadth of evidence, without resorting to additional (arbitrary) ingredients. As well as predicting the presence or absence of consciousness, such a theory can be tested on its predictions about the essential and accidental properties of experience in healthy adult humans, under circumstances where there should be little doubt about what the subject is experiencing.

While the "small-network" and "unfolding argument" remain debated, toy models have played a pivotal role in sharpening these discussions. They have highlighted the need for future work to focus on how various theories link experience to report, clarified the kinds of experiments that are admissible as test cases, and advanced our understanding of how a science of consciousness might develop when grounded in more nuanced notions of philosophy of science.

## Conclusion

As reviewed above, toy models may reveal much about the limits and possibilities of our theories of consciousness. In this context, Feynman's dictum—'What I cannot create, I do not understand'—serves as a guiding principle, not for building consciousness itself but for constructing and evaluating theories about it. By making abstract concepts tangible and accessible, toy models allow researchers to probe the coherence and consistency of theoretical frameworks. They distill complex phenomena into simplified systems, fostering a deeper understanding of proposed mechanisms and their implications. In this way, toy models offer intuitive insights that are often obscured in larger, more intricate systems or abstract formulations. At the same time, they serve as quantitative implementations of thought experiments, enabling researchers to rigorously test theoretical assumptions. Critically, toy models also help assess the scope of theories, determining whether their predictions are universal or constrained by their premises. For example, toy models have been pivotal in refining IIT's mathematical framework and illustrating its predictions, while also inviting criticism and exposing philosophical challenges. Additionally, toy models have highlighted the functional principles of GWT, while also drawing attention to the open question of what, within the theory, constitute necessary and sufficient conditions for consciousness.

While the primary focus of toy models has been to illuminate theoretical frameworks, they have also shown promise in addressing specific features of experience. For example, IIT's toy model of spatial extendedness provides a mechanistic account of a phenomenal property, linking it to the underlying cause-effect structure of a system. Moreover, toy models bring philosophical challenges into sharper focus. Debates surrounding the "small network" and "unfolding" arguments, for example, have underscored the broader challenges of relating conscious experience to physical systems. Toy models have also been used to argue that functional equivalence does not necessarily imply phenomenal equivalence, sharpening distinctions between structural and functional theories of consciousness.

Ultimately, toy models act as bridges between abstract theoretical claims and empirical science, providing invaluable tools for navigating the complexities of consciousness science. They elucidate not only the theories themselves but also the broader challenges that a comprehensive theory of consciousness must address. By engaging with toy models, scientists can refine their frameworks, tackle deep philosophical questions, and move closer to the ultimate goal of understanding consciousness in all its forms and manifestations.